\documentclass[a4paper]{article}

\usepackage{INTERSPEECH2020}

\usepackage{amsmath,amssymb}
\usepackage{url}
\usepackage{comment}
\usepackage{cleveref}
\DeclareMathOperator{\E}{\mathbb{E}}

\title{VQVC+: One-Shot Voice Conversion \\ by Vector Quantization and U-Net architecture}
\name{Da-Yi Wu$^1$, Yen-Hao Chen$^1$, Hung-Yi Lee$^1$}
\address{
  $^1$College of Electrical Engineering and Computer Science, National Taiwan University
\email{\{r07922119, r07921112, ,hungyilee\} @ntu.edu.tw }}

\begin{document}

\maketitle
\begin{abstract}
Voice conversion (VC) is a task that transforms the source speaker's timbre, accent, and tones in audio into another one's while preserving the linguistic content. 
It is still a challenging work, especially in a one-shot setting.
Auto-encoder-based VC methods disentangle the speaker and the content in input speech without given the speaker's identity, so these methods can further generalize to unseen speakers. 
The disentangle capability is achieved by vector quantization (VQ), adversarial training, or instance normalization (IN).
However, the imperfect disentanglement may harm the quality of output speech. 
In this work, to further improve audio quality, we use the U-Net architecture within an auto-encoder-based VC system. 
We find that to leverage the U-Net architecture, a strong information bottleneck is necessary. 
The VQ-based method, which quantizes the latent vectors, can serve the purpose. 
The objective and the subjective evaluations show that the proposed method performs well in both audio naturalness and speaker similarity.
   
\end{abstract}
\noindent\textbf{Index Terms}: voice conversion, vector quantization, skip-connection, disentangled representations

\section{Introduction}

Voice Conversion (VC) task is to convert the voice of the source speaker to the voice of a target speaker without losing the linguistic information in source speech. 
To imitate the target speaker, a VC system should modify the tone, accent, and vocalization of the source speaker. 
In tradition, VC focuses on one-one or many-to-one speaker transformation with parallel data, and it is treated as a statistical problem\cite{stylianou1998continuous, toda2007voice, helander2010voice}. 
However, the above methods need parallel data, which is challenging to collect.

Recently, many-to-many unparalleled VC has been studied. 
Generative adversarial network (GAN)\cite{goodfellow2014generative} and its variants, like cycleGAN \cite{kaneko2018cyclegan} and starGAN \cite{kameoka2018stargan}, are classic models to tackle unparalleled, many-to-many VC problems. 
Flow-based models like blow \cite{serra2019blow} have also been studied, and they transform waveforms directly instead of using acoustic features. 
These models directly covert the voice without feature disentangling behavior, achieving satisfactory audio quality.

 
 Other works~\cite{qian2020f0, lu2019one, andyliu2019, sxliu2018, lfsun2016, cchsu2016} attempt to disentangle the speaker's unit and content unit in the embedding space. These methods convert the voice by replacing the speaker embedding passed through the decoder.
  With a pre-trained speaker encoder, AutoVC~\cite{AutoVC} applies a vanilla auto-encoder and constrains the size of the latent representations to lead the encoder to extract the content information from the audio. 
  Chou et al.~\cite{AdaIN} bring the idea from the image style-transfer mission, referring the speaker to the style of speech; it maps different speakers to different values of mean and standard deviation, and utilizes AdaIN~\cite{huang2017arbitrary} to transform the speaker in speech. 
  VQVC~\cite{VQVC} applies vector quantization~\cite{van2017neural} technique to extract the content information, and learn to represent the speaker information by the difference between continuous space and the discrete codes. 
 Due to these models' ability to disentangle latent space, these approaches can synthesize the voice of the unseen speakers, and can even achieve one-shot VC, that is, the model synthesizes converted audio using only one sentence of each source and target speaker during the inference phase. 

To disentangle latent space, the model usually needs a strong bottleneck, which is constrained by GAN~\cite{chouGAN}, layer dimension~\cite{AutoVC}, IN~\cite{AdaIN}, or VQ~\cite{VQVC}, and due to the constraints applied on the model, the audio quality is sacrificed.
To deal with the above issue, we propose VQVC+, a U-Net architecture combining VQ and IN. 
U-Net connects each encoder layer and its corresponding decoder, and it has recently shown superior performance as a spectrogram generator~\cite{vasquez2019melnet,karras2017progressive, liu2019score}. 
However, U-Net~\cite{ronneberger2015u} is seldom used in voice conversion, and the reason is that its reconstruction capability is too well to make the model loss the ability to disentangle. 
In this paper, we find out that VQ can form a strong information bottleneck to prevent U-Net from overfitting on the reconstruction task, and with U-Net architecture, VQVC+ can hence synthesize high-quality audio. 
Also, we compare our method with AutoVC\cite{AutoVC} and Chou\cite{AdaIN}. 
The subjective evaluations show that the proposed method reaches a state-of-the-art result in one-shot VC. 
\section{Methods}
\subsection{VQVC}
VQVC~\cite{VQVC} is an one-shot voice conversion system with self-reconstruction loss. 
The core idea is: \textit{the content information can be represented by discrete codes}~\cite{chorowski2019unsupervised,liu2019towards}, \textit{and the speaker information can be viewed as the difference between the continuous representations and the discrete codes}.

\begin{figure}[t]
  \centering
  \centerline{\includegraphics[width=7.0cm]{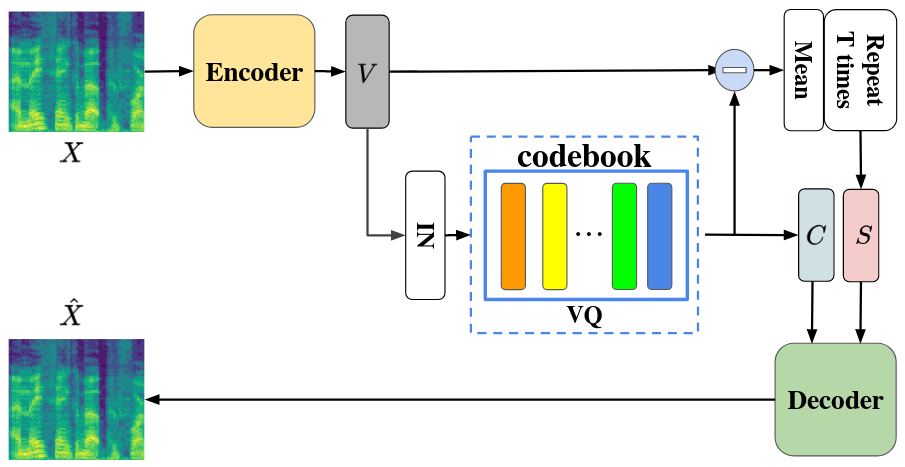}}
\caption{The VQVC architecture. VQ is the vector quantization layer, and IN is the instance normalization layer. VQVC applies IN+VQ layers to separate the content and the speaker information to achieve voice conversion.}
\label{fig:vqvc}
\end{figure}

As shown in \cref{fig:vqvc}, an auto-encoder architecture is used. $\mathcal{X}$ is our whole training set. 
We denote $\Vec X \in \mathcal{X}$ as an audio segment, represented as a sequence of acoustic features, $\Vec X=\{\Vec x_0, \Vec x_1, ...,  \Vec x_T\}$, where $T$ denotes the audio duration.
We denote $enc$ as the encoder, $dec$ as the decoder, $\mathcal{Q}$ as the quantization codebook, and $Quantize$ as the quantization functoin. 
Given the audio segment $\Vec X$, the continuous latent representation $\Vec V \in R^{F \times T}$, the content embedding $\Vec C \in R^{F \times T}$, and the speaker embedding $\Vec S \in R^{F \times T}$ can be derived as 
\begin{equation}
\vspace{-8mm}
\label{eq:emb}
\begin{aligned}
\Vec V &= enc(\Vec X),\\
\Vec C &= Quantize(\Vec V), \\
\Vec s &= \E_t [\Vec V - \Vec C], \quad \Vec S = \{\underbrace{\Vec s, \Vec s, ..., \Vec s}_{T\ times}\},
\end{aligned}
\end{equation}
\\
\\
where $F$ is embedding size, and
\begin{equation*}
Quantize(\Vec V) = \{\Vec q_0, \Vec q_1, ..., \Vec q_T\}, \quad \Vec q_j = \arg\min_{\Vec q \in \mathcal{Q}}(||\Vec v_j - \Vec q||_2^2).
\end{equation*}
Note that the dimension of any vector $\Vec v_j$ is equal to the dimension of a code in $\mathcal Q$. 
The instance normalization (IN)~\cite{ulyanov2016instance} layer is added before the quantization, which is indispensable for good performance.
We expect that the speaker information is a kind of global information of the speech. Hence, we derive $\Vec s$ by subtracting $\Vec C$ from $\Vec V$ and then take expectation $\E_t$ on the utterance duration, representing the global information of the audio segment. Afterwards, we get $\Vec S$ by repeating $\Vec s$ for T times and concatenate them to enforce the dimensions of $\Vec C$ and $\Vec S$ to be the same. 
Then, $\Vec S$ is added back to $\Vec C$, passed through the decoder, and at last we get the reconstruction
\begin{equation}
    \hat{\Vec{X}} = dec(\Vec C+\Vec S).
\end{equation}

In the training phase, the reconstruction loss can be written as \Cref{eq:loss_rec}: 
\begin{equation}
\label{eq:loss_rec}
    L_{rec}(\mathcal{Q},\theta_{enc},\theta_{dec}) = \E_{\Vec X \in \mathcal{X}}\big[||\hat{\Vec X} - \Vec X||_1^1\big].
\end{equation}
In addition, the latent loss $L_{latent}$ is added as \Cref{eq:loss_latent}, which minimizes the distance between the discrete codes and the  continuous embedding. We denote $IN$ as instance normalization layer.
\begin{equation}
\label{eq:loss_latent}
    L_{latent}(\theta_{enc}) = \E_t[||IN(\Vec V) - \Vec C||^2_2]. \\
\end{equation}
The whole loss can be written as \Cref{eq:loss}:
\begin{equation}
\label{eq:loss}
    L = L_{rec} + \lambda L_{latent}.
\end{equation}
During the inference phase, the content embedding $\Vec C$, and the speaker embedding $\Vec S$ would be extracted from different speakers.


\begin{figure}
  \centering
  \centerline{\includegraphics[width=7.0cm]{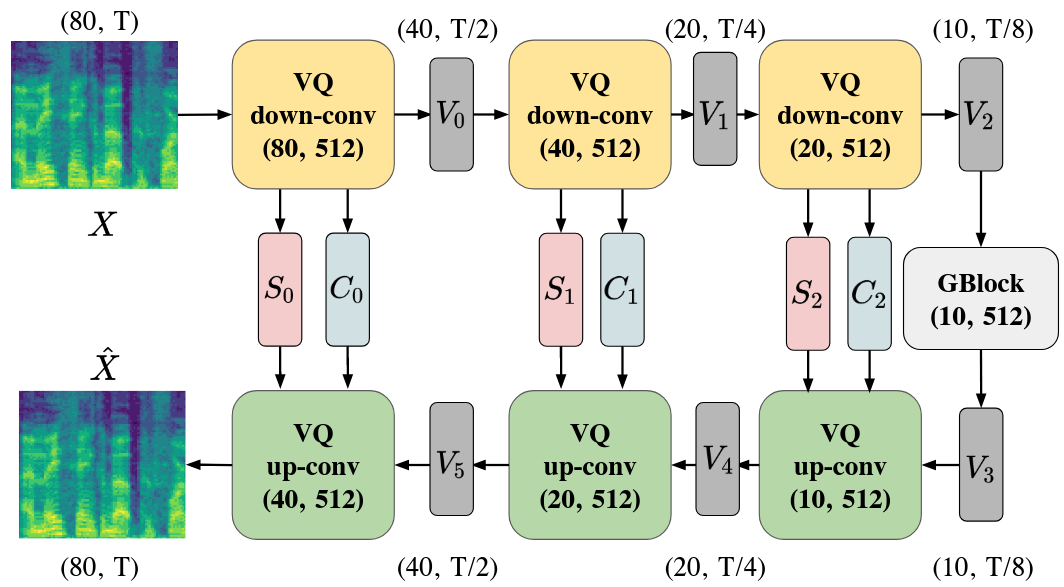}}
\caption{The VQVC+ architecture. VQVC+ applies the U-Net architecture to improve quality, and each sub-module in the encoder is a variant of the VQVC encoder. Quantized output $\Vec C$  and the speaker embedding $\Vec S$ are skip-connected to the decoder instead of the continuous embedding $\Vec V$.}
\label{fig:vqvc+}
\end{figure}

\subsection{VQVC+}
Although VQVC can disentangle the linguistic content and the speaker information well, the synthesized audio quality still has room for improvement.
VQVC synthesizes the audio that matches the target speaker's characteristic, but the vocalization of the audio is vague. 
We attribute the problem to the information loss induced by the vector quantization, which makes the decoder unable to reconstruct the content properly.  
Thus, to improve synthesis quality, we apply the U-Net architecture, which has shown superior performance as a spectrogram generator~\cite{ronneberger2015u}, on VQVC, and we call this new models VQVC+. 
 
\Cref{fig:vqvc+} illustrates the whole architecture of VQVC+. The encoder is composed of three $VQ\text{ }down\text{-}conv$ modules, which is a variant of VQVC encoder; the decoder is composed of three $VQ\text{ }up\text{-}conv$ modules. To strength the content information received by the decoder layers, the content embedding $C$ and the speaker embedding $S$ are skip-connected into the corresponding decoder layer, which is in a similar light with U-Net. 


\subsubsection{VQ down-conv module}
\begin{figure}[t]
  \centering
  \centerline{\includegraphics[width=7.0cm]{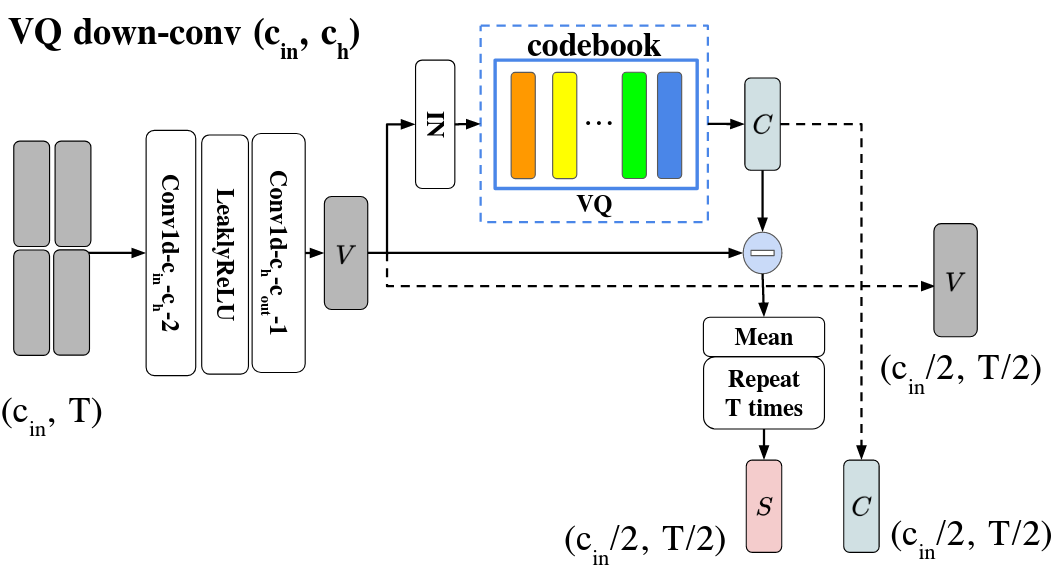}}
\caption{VQ down-conv module. The Conv1d-c$_1$-c$_2$-N indicates a 1D-convolution layer with input channel c$_1$, output channel $c_2$, and stride N.}
\label{fig:downconv}
\end{figure}

As shown in \Cref{fig:downconv}, $VQ\text{ }down\text{-}conv$ is comprise of two $3 \times 1 $ kernel 1D-convolution layers, an IN layer, and a vector quantization layer. The $Conv1d\text{-} c_1\text{-}c_2\text{-}N$ layer indicates the 1D-convolution layer, whose input channel and the output channel are $c_1$ and $c_2$ respectively, and $N$ denotes the stride. 

$VQ\text{ }down\text{-}conv(c_{in}, c_h)$ takes a matrix with dimension $(c_{in}, T)$ as input and outputs three components, $\Vec V$, $\Vec C$, and $\Vec S$. $\Vec V$ is the embedding of continuous space directly get from the convolution block; $\Vec C$ is the quantized matrix of $\Vec V$ passing through IN and VQ; $\Vec S$ is the speaker embedding as we mentioned in \Cref{eq:emb}. The dimensions of $\Vec V$, $\Vec C$, and $\Vec S$ are $(c_{in}/2, T/2)$, $(c_{in}/2, T/2)$, and $(c_{in}/2, T/2)$ respectively.  

\subsubsection{VQ up-conv module}
\begin{figure}[t]
  \centering
  \centerline{\includegraphics[width=7.0cm]{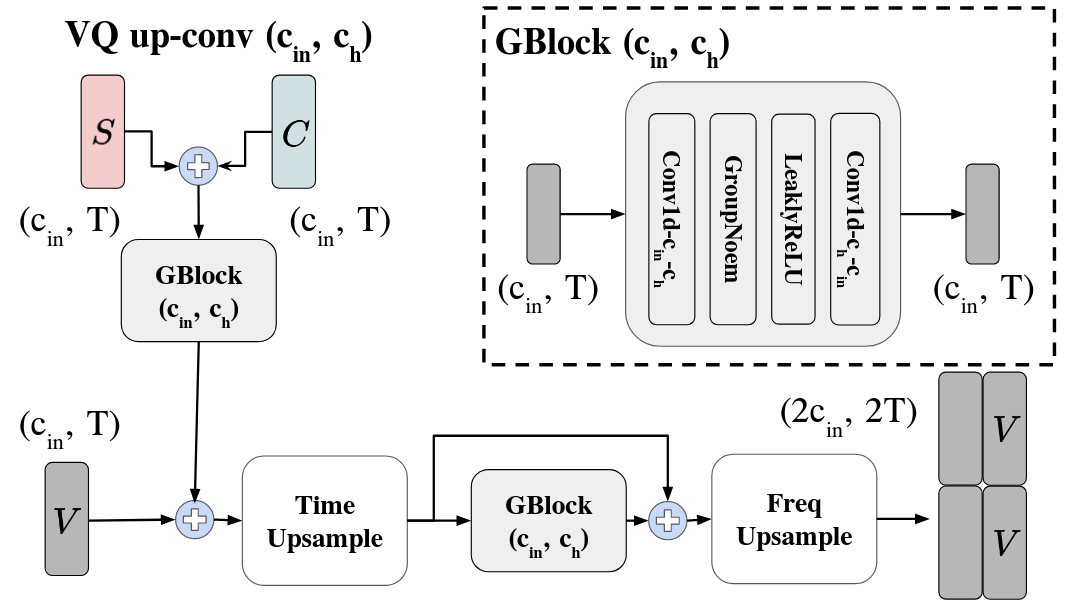}}
\caption{VQ up-conv module and Group Norm Block (GBlock)}
\label{fig:upconv}
\end{figure}

As shown in \Cref{fig:upconv}, $VQ\text{ }up\text{-}conv$ takes the output of previous layer, $\Vec V$, and $\Vec C$, $\Vec S$ generated from the corresponding encoder layer as inputs. 
Embeddings are added and upsampled by a factor of 2 within both frequency and time domain. $VQ\text{ }up\text{-}conv$ contains three main components, Group Norm Block (GBlock)~\cite{liu2019score}, TimeUpsampling, and FreqUpsampling. 

GBlock consists of two $3 \times 1 $ kernel 1D-convolution layers and groupnorm following with LeakyReLU, where the size of its input and output are the same. 

\begin{figure}[t]
  \centering
  \centerline{\includegraphics[width=8.0cm]{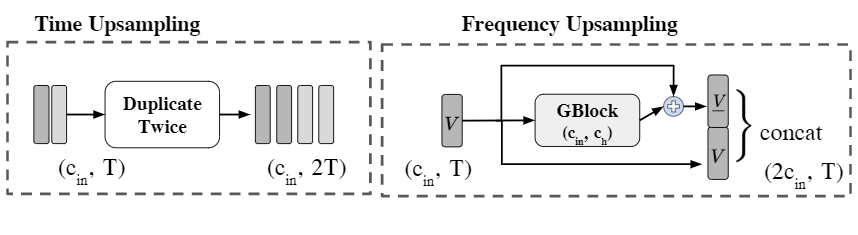}}
\caption{Upsampling block}
\label{fig:upsample}
\end{figure}

As shown in \Cref{fig:upsample}, the TimeUpsampling module duplicates each vector twice to expand the time dimension; the FreqUpsampling module emphasizes the importance of low-frequency area in mel-spectrogram, so it generates the high-frequency part using its low-frequency part and concatenates them as its output.

$\Vec C$ and $\Vec S$ are first added and pass through the GBlock, and we add $\Vec V$ from the previous layer on it afterward. Then, it passes through the two upsampling modules to get the output.

\subsubsection{U-Net}
Our architecture can be seen as a variant of U-Net. As shown in \Cref{fig:vqvc+}, each $VQ\text{ }down\text{-}conv$ module generates its own $\Vec V$, $\Vec C$, $\Vec S$, where $\Vec V$ is passed through the next $VQ\text{ }down\text{-}conv$ module, and $\Vec C$, $\Vec S$ are passed through the corresponding $VQ\text{ }up\text{-}conv$ module in the decoder. 
Model is trained with latent loss $L_{latent}$ for each layer and reconstruction loss $L_{rec}$. 
We assign equal weight $\lambda$ in (\ref{eq:loss}) for the latent loss $L_{latent}$ of all layers during training.

\section{Experimental Setup}

\subsection{Datasets}
\label{Datasets}
We conduct our experiment on the VCTK dataset\cite{veaux2016superseded}, which contains about 46 hours of audio from 109 speakers, and there are about 500 sentences for each speaker. We select 20 speakers as our testing set, where we denote them as our unseen speakers. 
For each audio, we remove the silence and randomly choose 3 seconds for training. If the duration of the audio is less than 3 seconds, we repeat it into 3 seconds. Then, we convert the audio from 48000Hz into 22050Hz and perform STFT with 1024 STFT window size and 12.5 milliseconds hop size. Next, we transform the magnitude of the spectrograms to 80-bin mel-scale and take logarithm. 
To convert the mel-spectrogram back to the waveform, we apply the fast, high-quality pre-trained MelGAN vocoder\cite{melgan}.

\subsection{Training details}
We train the proposed model using ADAM optimizer with a 0.01 learning rate, and $\beta_1$ = 0.9, $\beta_2$ = 0.999. Our channel size in each $VQ\text{ }down\text{-}conv$ and $VQ\text{ }up\text{-}conv$ is 512, and our codebook size in each encoder layer is 64. We set the batch size to 32, and latent loss factor in E.q.4, $\lambda$, to 0.1. We train our model on 99 speakers for 200k iterations. Further details may be found in our implementation code\footnote{\url{https://github.com/ericwudayi/SkipVQVC}}.

\section{Experiments}
\subsection{Content embedding}
\label{aod}
\begin{figure}[t]
  \centering
  \centerline{\includegraphics[width=8.0cm]{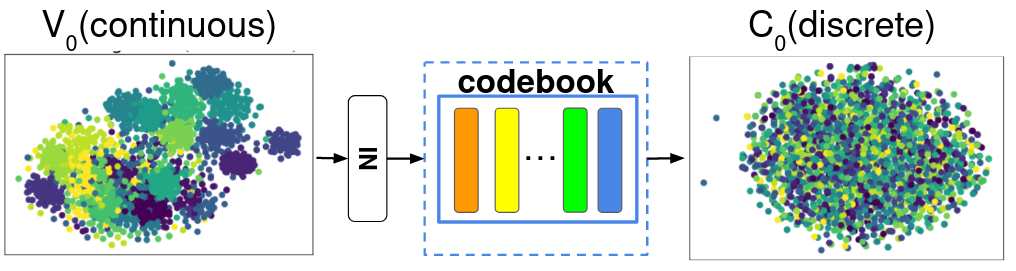}}
\caption{quantization}
\label{fig:quantization}
\end{figure}

We first demonstrate the effects of IN and VQ in \Cref{fig:quantization}. 
We perform t-SNE on $\Vec V_0$ and $\Vec C_0$ of 20 distinct speakers.
Different colors represent different speakers. 
It is shown that for $\Vec V_0$, the points with the same colors tend to be clustered, whereas there is no obvious group in $\Vec C_0$.

To further verify that the speaker information is discarded by the IN and VQ layers, we train a speaker classifier based on content embedding.
The lower speaker classifier accuracy, the less speaker information contains in content embedding.
The classifier is composed of three 1D-convolution layers with 256 hidden nodes followed by a fully connected layer. 
We compare the result with different quantization settings. Here, Q32(64, 128, 256) represents VQVC+ that the codebook size of each encoder layer is 32(64, 128, 256); IN-only means there is no quantization module after the IN layers; VQVC~\cite{VQVC} is the original model without skip-connection, and its codebook size is 128. 

We conduct our experiment on the output of each encoder layer, $\Vec C_0$, $\Vec C_1$, and $\Vec C_2$. As shown in \Cref{tab:quantize}, deeper layers have lower accuracy for every model, and the accuracy of the VQ models are apparently lower than IN-only. IN-only has 71.2\% speaker identification rate on $\Vec C_0$, which indicates that IN-only has no ability to disentangle the content information and the speaker information. 
We listen to the audio synthesized using IN-only, finding that most of them cannot perform conversion; IN-only just reconstructs the source audio. 
The codebook size of a model determines how strong the information bottleneck is. Models with smaller codebook size, like Q32, can achieve lower speaker identification rate, but it may lose more content information, which makes higher reconstruction error. On the other hand, models with larger codebook size, like Q256, can reconstruct the audio well, but it may leak some speaker information to its quantized code. 
We choose Q64 in the following experiments, which achieves an exceptional balance on reconstruction and disentanglement. 

\begin{table}[h] 
 \centering
 \begin{tabular}{||c| c| c||} 
 \hline
 Method &  $\Vec C_0$ /$\Vec C_1$ / $\Vec C_2$ (\%) & L1Loss\\ [0.5ex] 
 \hline
 VQVC & 16.0  &0.262 \\
 \hline
 Q32 & 19.5 / 11.8 / 6.8  &0.210\\ 
 \hline
 Q64 &  23.2 / 16.6 / 7.0 &0.188\\
 \hline
 Q128 & 33.3 / 17.0 / 10.3  & 0.180\\
 \hline
 Q256 & 35.8 / 18.1 / 12.5  &  0.165\\ 
 \hline
 IN-only & 71.2 / 36.8 / 5  &  0.145\\
 \hline
\end{tabular}
\caption{Accuracy of identifying speakers on the content embedding and the speaker embedding with different methods. VQVC is the model without skip-connection design. Q$N$ means that the size of codebook, $Q$, in VQVC+ is $N$. IN-only means no quantization in U-Net. L1Loss is the L1 reconstruction loss.}
\label{tab:quantize}
\end{table}
\subsection{Speaker embedding}

We do not explicitly add any speaker-relative objective or constraint to the encoder, while the speaker embedding is learned properly due to the effective disentanglement of the quantized content embedding.
To show that our extraction-based speaker embedding is speaker-correlated, we select 20 unseen speakers to generate their speaker embeddings. 
For each of $\Vec S_0$, $\Vec S_1$, and $\Vec S_2$, we train a classifier to identify which speaker is it, and the classifier's architecture is the same as we mentioned in \Cref{aod}.
The higher the classifier accuracy, the better the speaker embedding is.
The model used in this experiment is Q64, which is the same one mentioned in \Cref{aod}. The results are shown in  \Cref{tab:quantize}. It presents that the speaker information in $\Vec S_0$ is learned very well. 
$\Vec S_1$ and $\Vec S_2$ have 72.2\% and 45.4\% accuracy, which indicates that the model extracts the speaker embeddings at lower resolution spaces.

\begin{table}
 \centering
 \begin{tabular}{||c| c |c||} 
 \hline
 Method & $\Vec S_0/ \Vec S_1/ \Vec S_2 $ (\%)\\ [0.5ex] 
 \hline
 VQVC &  96.6\\
 \hline
 Q64 &  98.3 / 72.2/ 45.4  \\
 \hline
 IN-only & 97.4/ 80.1 / 23.1   \\
 \hline
\end{tabular}
\caption{Accuracy of identifying speakers on the speaker embedding $\Vec S$.}
\label{tab:speaker_embed}
\end{table}

\begin{figure}[t]
  \centering
  \centerline{\includegraphics[width=6.0cm]{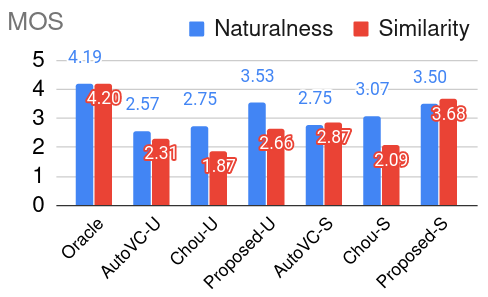}}
\caption{MOS of different methods. "-U" and "-S" are refer to unseen speakers and seen speakers respectively.}
\label{fig:mos}
\end{figure}

\begin{figure}[ht!]
  \centering
  \centerline{\includegraphics[width=6.0cm]{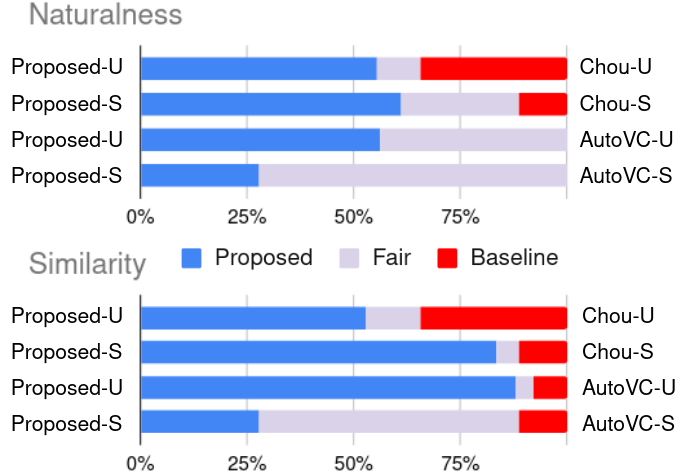}}
\caption{Pairwise comparison.  For each row, the blue bar represents the percentage that the proposed method is preferred, and the red bar for the baseline. "-U" and "-S" mean unseen speakers and seen speakers respectively.}
\label{fig:pairwise}
\end{figure}

\vspace{-2mm}
\subsection{Subjective evaluations}
\label{ssec:sev}

We conduct subjective tests to evaluate the converted audio's sound quality and its speaker similarity to the target one. We choose AutoVC\cite{AutoVC} and Chou\cite{AdaIN} as our baseline models, and we train the models using the official code\footnote{\url{https://github.com/auspicious3000/autovc}}\footnote{\url{https://github.com/jjery2243542/adaptive_voice_conversion}} by ourselves. Note that the training data of AutoVC(Chou) has 40(20) speakers in their original paper implementation. While in our experiments,  for a fair comparison, all these models are trained with the same 99 speakers, and the output mel-spectrograms are converted to waveform using a pretrained MelGAN vocoder\cite{melgan}.

\Cref{fig:mos} shows the results of the mean opinion score (MOS) test. The "-S" represents that the speakers of the source and the target audio are in the training set, which means they are seen speakers. Otherwise, the "-U" denotes that the speakers are unseen. We ask the subjects (1) how natural and (2) how similar to the target speaker the converted audio sounds, and they score from 1 (very bad) to 5 (very good) after listening to the converted audio and the target audio. "Oracle" means that the audio is generated from MelGAN vocoder with real mel-spectrograms. Hence, it becomes the upper-bound of these three models. AutoVC uses pre-trained speaker embedding, while Chou and our proposed method do not. The results show that our speaker embedding is more effective in restoring speaker information for VC. Moreover, our proposed method also performs the best in the result of naturalness, which indicates that our model generates better mel-spectrograms than other baselines.

Except for the MOS test, we also perform pairwise comparisons between our proposed method and the other two baseline methods. For each question within the questionnaire, we sample audio generated from our model and the baseline model in random order, and ask the subjects which audio sounds more natural and more similar to the target speaker. As shown in \Cref{fig:pairwise}, our proposed method beat the others in both seen("-S") and unseen("-U") senarios. Further, we observe that in seen scenarios("-S"), AutoVC and our method are comparable. Nevertheless, in unseen("-U") scenarios, our method wins for most cases. This implies that our model is more robust to unseen speakers. Meanwhile, it indicates that our method apparently performs better than the baselines for one-shot VC. The generated audio can be found in our demo page.\footnote{\url{https://ericwudayi.github.io/VQVC-DEMO}}

\section{Conclusions}
In this paper, we present a new model for one-shot VC. We use the U-Net combined with VQ layers to achieve a high-quality VC. With the well-designed architecture, our proposed model is able to separate the speaker information and the content information effectively in an elegant way with the self-reconstruction loss only. The objective results verify the strong disentanglement of our model, while the subjective results can support our conjecture that the skip-connection design is beneficial for achieving high-quality conversion. 



\bibliographystyle{IEEEtran}

\bibliography{mybib}


\end{document}